\documentstyle[11pt,newpasp,twoside,epsf]{article}
\begin{document}
\title{Molecular Gas \& Star Formation in Nearby Galaxies}
\author{Tony Wong}
\affil{CSIRO Australia Telescope National Facility, PO Box 76, Epping
NSW 1710, Australia; and School of Physics, University of New South
Wales, Sydney NSW 2052, Australia}
\author{Michele D. Thornley}
\affil{Department of Physics, Bucknell University, Lewisburg, PA
17837, USA}

\begin{abstract}
We review recent observations of molecular gas in nearby galaxies and
their implications for the star formation law on large ($>1$ kpc)
scales.  High-resolution data provided by millimetre interferometers
are now adding to the basic understanding that has been provided by
single-dish telescopes.  In particular, they confirm the good
correlation between star formation rate (SFR) and molecular gas
surface densities, while at the same time revealing a greater degree
of heterogeneity in the CO distribution.  Galaxies classified as SAB
or SB tend to show radial CO profiles that peak sharply in the
inner $\sim$20\arcsec, indicative of bar-driven inflow.  The observed
Schmidt law index of $\approx$1.5 may result from a nearly linear
relation between SFR and H$_2$ mass coupled with a modest dependence
of the molecular gas fraction on the total gas density.  The
normalisation of the Schmidt law, giving the characteristic timescale
for star formation, may stem from the generic nature of interstellar
turbulence.
\end{abstract}

\section{Observational Tracers}

The principal way to trace molecular gas in galaxies is with the
rotational lines of the CO molecule, due to CO's relatively high
abundance (about $10^{-5}$ of H$_2$) and low excitation requirements
($\Delta E_{1\rightarrow0}/k = 5.5$ K).  The high abundance of CO generally
makes it optically thick, especially in the lowest ($1\rightarrow0$)
transition, with the result that the {\it effective} critical density
is quite low, $n_{cr}/\tau \sim 300$ cm$^{-3}$, and the excitation
temperature $T_{ex}$ approaches the kinetic temperature in most
molecule-rich regions.  Unlike CO, H$_2$ itself has no
permitted rotational transitions, and can only be observed in infrared
(quadrupolar or rovibrational) lines that require more extreme
($\Delta E/k > 500$ K) excitation conditions.

The main disadvantage of CO is that its high opacity makes it a poor
tracer of column density.  Indeed, the theoretical basis for using CO
as a column density tracer is the {\it virial hypothesis}: the CO
emission comes from an ensemble of virialised clouds that do not
shield each other in position-velocity space.  For a virialised cloud,
the size $R$, density $\rho$, and linewidth $\Delta v$ are related by:
\begin{equation}
\Delta v \propto \rho^{1/2} R\;.
\end{equation}
Then the CO luminosity is proportional to the H$_2$ mass, assuming
roughly constant density and temperature:
\begin{equation}
L_{\rm CO} \propto T\Delta v R^2 \propto T\rho^{1/2} R^3 \propto
\frac{T}{\rho^{1/2}}M_{\rm H_2}\;.
\end{equation}
However, the assumption of virial equilibrium in molecular clouds has
been questioned (e.g.\ Ballesteros-Paredes \& Mac Low 2002), and in
regions where CO comes predominantly from a diffuse intercloud
medium---as has been suggested for starburst galaxies (Solomon et al.\
1997, Downes \& Solomon 1998)---the linearity between CO emission and
H$_2$ column density is likely to break down.  CO is also not as
self-shielding at H$_2$, and might be dissociated even in regions
where H$_2$ is present.  Finally, its abundance relative to H$_2$ will
vary according to the metallicity of the interstellar medium (ISM),
and so it is unlikely to be detectable in very low metallicity
regions.

Other methods to trace molecular gas include observations of
far-infrared and submillimetre emission from dust, and ultraviolet (UV)
H$_2$ absorption towards continuum sources.  Dust emission is usually
optically thin and depends only linearly on temperature in the
Rayleigh-Jeans part of the spectrum, although there are additional
uncertainties arising from the adopted grain size distribution and
gas-to-dust ratio.  Alton et al.\ (2002) show that the 850$\mu$m dust
emission from the disk on NGC 6946 matches the CO emission very well.
The UV absorption technique is biased against high gas columns because
of extinction, but does provide a powerful tool to study {\it diffuse}
H$_2$.  Tumlinson et al.\ (2002) find that the diffuse H$_2$ fraction
in the Magellanic Clouds is very low, $\sim$1\% compared to $\sim$10\%
in the Galaxy.

For the remainder of this paper, we assume that CO can be used as a
quantitative tracer of H$_2$ within the disks of normal spiral
galaxies.

\section{Radial Gas Profiles}

Our general knowledge of the CO distribution in galaxies is still
based on the single-dish survey of some 300 galaxies conducted
with the FCRAO 14-m telescope (Young et al.\ 1995).  Using
multiple-pointing observations for 193 of the galaxies, Young et al.\
concluded that CO is usually peaked toward galaxy centres: only in 28
galaxies was there an indication of a molecular ring or off-centre CO
peak.  At higher resolution, however, the picture is not so simple.
The BIMA Survey of Nearby Galaxies (BIMA SONG), which imaged the
distribution of CO emission in 44 nearby spirals at
6\arcsec--9\arcsec\ resolution and included single-dish data for full
flux recovery, finds that the CO distribution in spiral galaxies is
generally very heterogeneous (Regan et al.\ 2001, Helfer et
al.\ 2003).  Even when azimuthally averaged, large departures from a
smooth exponential profile are seen, due to the very clumpy
distribution of molecular gas (which responds strongly to spiral arms,
bar perturbations, etc.).  In addition, only 20 of the 44 SONG galaxies
exhibit their maximum CO surface brightness within the central
beamwidth, with six galaxies showing no detectable CO emission at all in
this region (Helfer et al.\ 2003).

\begin{figure}
\plotone{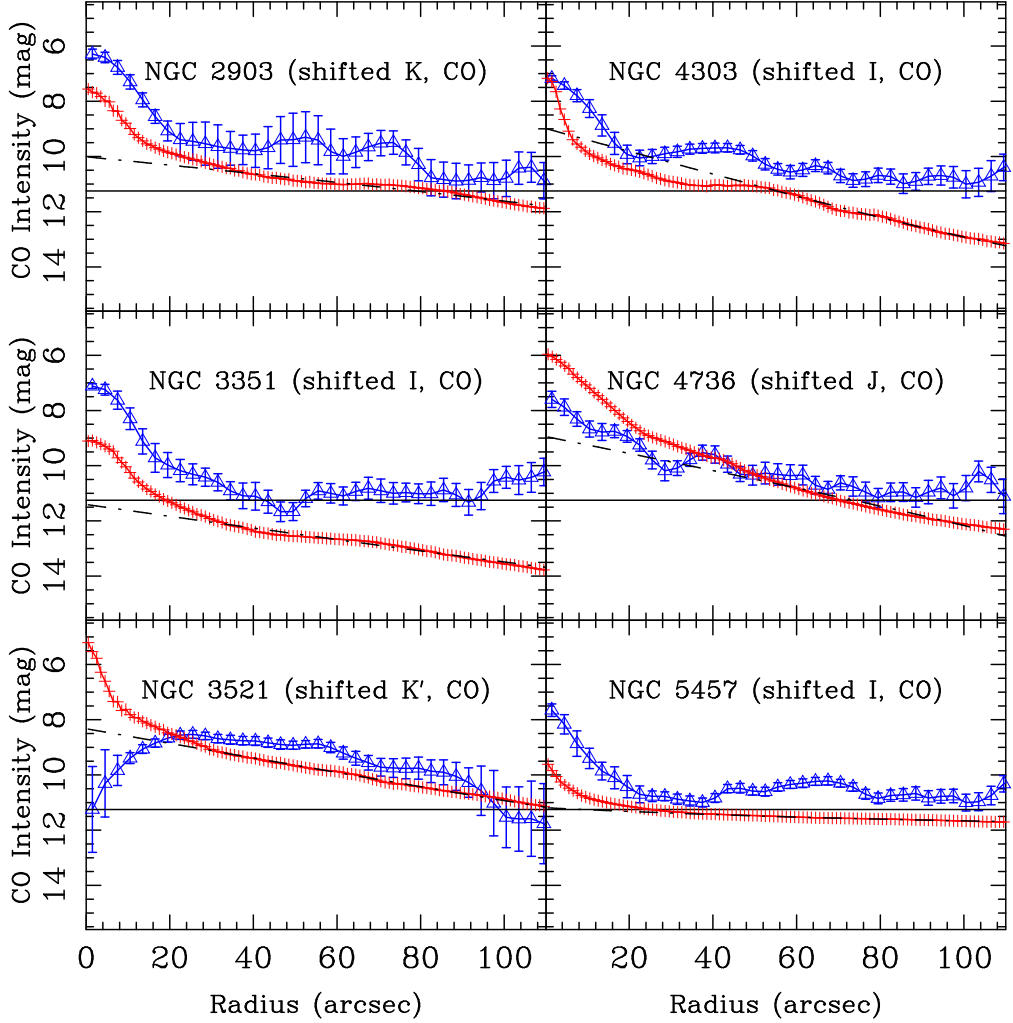}
\caption{Radial CO (triangles) and stellar (crosses) profiles for six
galaxies from the BIMA SONG sample (Thornley et al.\ 2003).  The
stellar profiles are taken from $I$, $J$, or $K$ band images as
indicated and are shifted vertically for comparison with the CO
profile.  The CO brightness is expressed in hyperbolic magnitude units
(Lupton et al.\ 1999) with 0 mag = 1000 Jy km s$^{-1}$ arcsec$^{-1}$.
The dot-dashed line indicates an exponential fit to the outer ($r >
50\arcsec$) part of the stellar profile, while the solid horizontal
line represents the magnitude consistent with no CO flux.}
\end{figure}

With the higher resolution imaging, the distributions of molecular
gas and stars can be compared on scales relevant for studying the
effects of internal dynamical processes.  Even though many
galaxies in BIMA SONG do not show central CO peaks, the majority
do display a CO excess in the inner $\sim$20\arcsec\ above the
exponential profile, where the stellar bulge contributes a similar
excess to the stellar light profile (Regan et al.\ 2001, Thornley et
al.\ 2003).  Representative profiles are shown in Figure 1.
Concentrating on the 27 BIMA SONG galaxies with the most extended CO
detections, Thornley et al.\ (2003) have found that such CO excesses
occur in both early and late-type galaxies, and are nearly universal
in galaxies with some bar contribution (type SAB or SB).  This
suggests that at least some of the central excesses are due to
bar-driven inflow of molecular gas (Sakamoto et al.\ 1999, Sheth et
al. 2004).  However, roughly half of the selected galaxies without
significant bar contributions also show such central excesses,
suggesting that significant bars are not required to produce gas
inflow.

It is notable that the HI radial distribution is almost always much
flatter than the CO, and often shows a central depression, as if the
atomic gas has undergone a phase transition to form H$_2$.  Recent
comparisons of single-dish CO maps and VLA HI imaging by Crosthwaite
et al.\ (2001, 2002) highlight this dichotomy.  Wong \& Blitz (2002)
found that for seven galaxies with high-resolution CO and HI data, the
HI/CO ratio increases with radius as roughly $R^{1.5}$, consistent
with being determined largely by the hydrostatic pressure of the ISM,
as predicted by Elmegreen (1993).  In this interpretation, star
formation is rarely found in low-pressure regions such as the halo or
outer disk because the dominant phase of neutral gas there is atomic.

\section{The Star Formation Law}

The FCRAO survey indicated that CO emission scales roughly linearly
with star formation tracers (e.g. Rownd \& Young 1999) except in
merging or interacting galaxies (Young et al.\ 1996), which show
enhanced star formation rates.  Kennicutt (1998), averaging CO, HI,
and H$\alpha$ fluxes within the optical disks of 61 galaxies, found a
strong correlation between the SFR and the total gas content,
consistent with a Schmidt (1959) law:
\begin{equation}\label{obeqn}
\Sigma_{\rm SFR} = 0.16\, \Sigma_{\rm gas}^{1.4}\;,
\end{equation}
where $\Sigma_{\rm SFR}$ and $\Sigma_{\rm gas}$ are the surface
densities of the star formation rate (in $\rm
M_\odot\;pc^{-2}\;Gyr^{-1}$) and total gas mass (in $\rm
M_\odot\;pc^{-2}$) respectively.  The correlation of SFR with HI or CO
individually was much poorer, suggesting that the gas involved
in star formation can quickly cycle between atomic and molecular
phases.

How does the star formation law behave on smaller scales?  Wong \&
Blitz (2002) compared CO, HI, and H$\alpha$ emission in seven
galaxies from the BIMA SONG sample, and found that the
Schmidt law continues to hold for azimuthally averaged rings spaced by
$\sim$1 kpc.  The exact slope of the power law depends on how one chooses to
correct for H$\alpha$ extinction; an extinction-free SFR indicator,
such as the total IR emission (Kewley et al.\ 2002) might yield a more
accurate slope than H$\alpha$.  Regardless, the correlation of SFR with CO is
considerably stronger than that with HI, as the HI profile is often
declining or flat in the inner regions, whereas star formation is
generally centrally peaked (Figure 2).  It is unclear whether the poor
correlation of SFR with CO found by Kennicutt (1998) may be due to
differences in galaxy selection, as the galaxies studied by Wong \& Blitz
(2002) are relatively strong CO emitters.

\begin{figure}
\plottwo{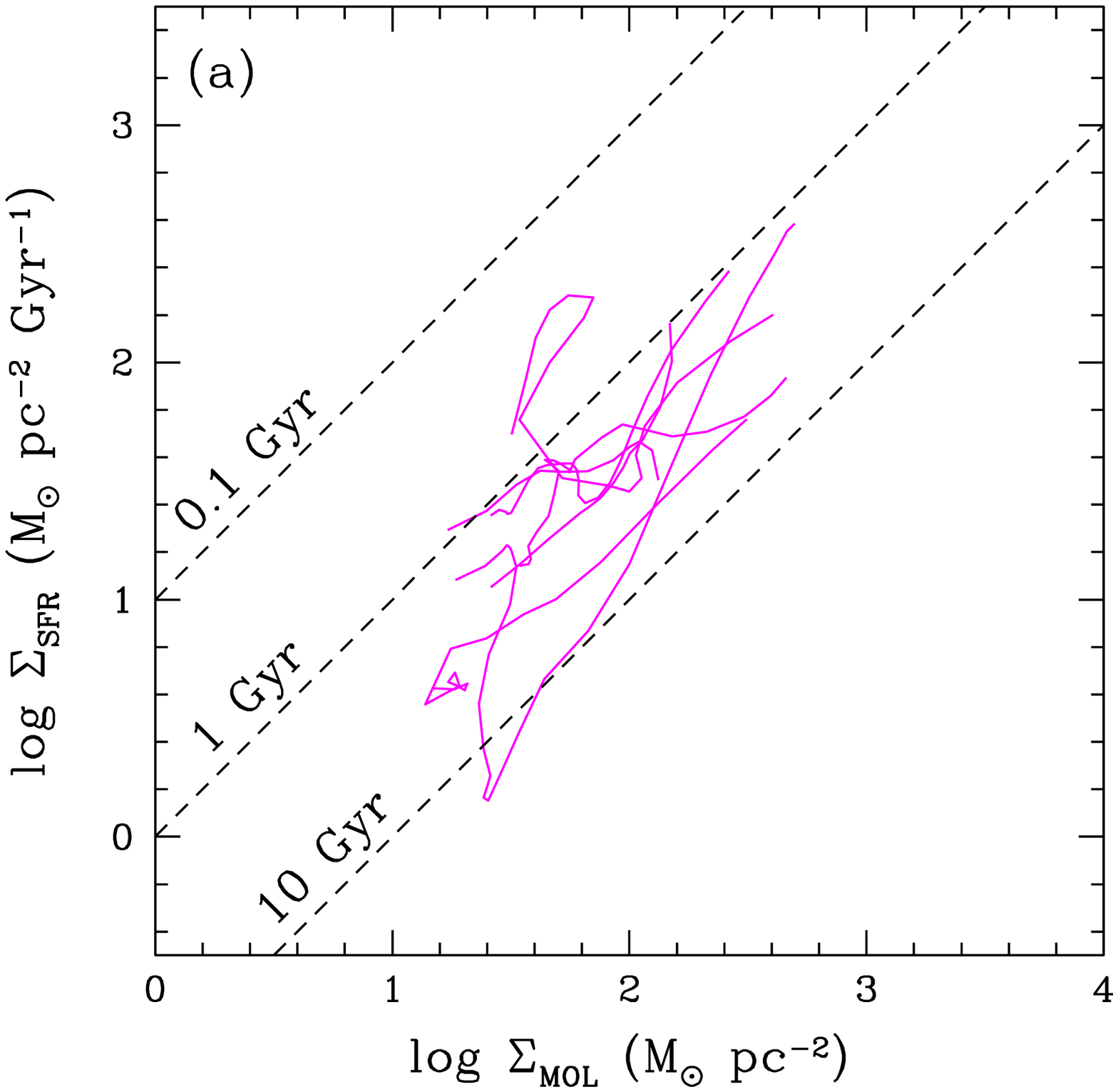}{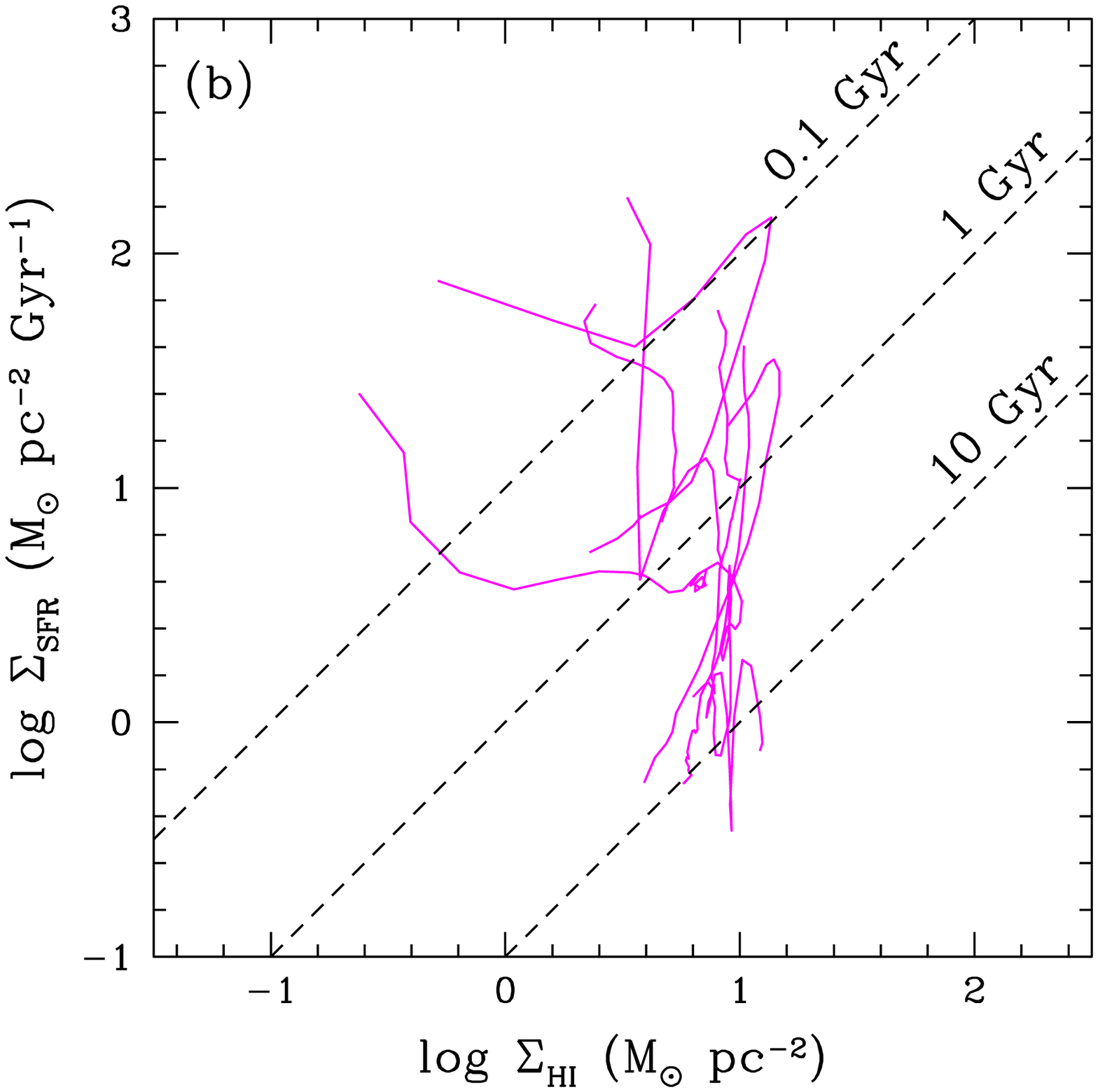}
\caption{SFR surface density plotted against the surface density of
(a) H$_2$ (b) HI, for seven galaxies studied by Wong \& Blitz (2002).
Each line represents azimuthally averaged data for a range of radii in
a single galaxy.  Dashed lines represent lines of constant gas
depletion time.  The spatial resolution of the images is about
6\arcsec\ for panel (a) and 15\arcsec\ for panel (b).}
\end{figure}

If SFR generally correlates with molecular rather than atomic gas,
then the Schmidt law index of $\approx$1.5 may derive from a
roughly linear relation between SFR and H$_2$ coupled with a weak
dependence of the molecular fraction $f_{\rm mol}$ on $\rho_{\rm
gas}$:
\begin{equation}
\rho_{\rm SFR} \propto \rho_{\rm mol}\,,\quad f_{\rm mol} \propto
\rho_{\rm gas}^{0.5}\;.
\end{equation}
This implies that star formation is a two-step process involving
molecular cloud formation, and contrasts with the conventional
interpretation that the index results from the free-fall (dynamical)
time being the natural timescale for star formation:
\begin{equation}\label{theqn}
\rho_{\rm SFR} = \frac{\epsilon\rho_{\rm gas}}{(G\rho_{\rm
gas})^{-0.5}} \propto \rho_{\rm gas}^{1.5}\;.
\end{equation}
Indeed, it has long been known that the star formation timescale is
much longer than the free-fall time; this inefficiency of star
formation is reflected in the numerical factor in Eq.~\ref{obeqn} and
$\epsilon$ in Eq.~\ref{theqn}.  Recently, two studies have tried
to explain the observed Schmidt law using the density probability
distribution function (PDF) that results from interstellar turbulence,
as revealed in simulations by (e.g.) Wada \& Norman (2001).
Elmegreen (2002) hypothesises that only that part of the density PDF
above $10^5$ cm$^{-3}$ proceeds to form stars on a dynamical
timescale, whereas Kravtsov (2003) assumes a much lower critical
density of 50 cm$^{-3}$ but a much longer star formation timescale of
4 Gyr.  In both cases, the presence of a critical density ensures that
only a small fraction of the galactic gas is involved in star
formation at a given time.  Detailed case studies of nearby galaxies
such as M33 and the LMC may be able to shed further light on this
issue.

\acknowledgments
We thank Eva Schinnerer for providing the IRAM PdB image of NGC
4736 used in the talk, and Leo Blitz for extensive discussions over
the past few years on many of the topics covered here.  This work was
supported by a Bolton Fellowship awarded to T.W. by the CSIRO ATNF.

\end{document}